\documentstyle[aps,prc,preprint,epsf,tighten]{revtex}
\begin{document}
\title{Electromagnetic Meson Form Factors in the Salpeter Model}

\author{C.R.\ M\"unz, J.\ Resag, B.C.\ Metsch, H.R.\ Petry}
\address{Institut f\"ur Theoretische Kernphysik,\\
         Universit\"at Bonn, Nussallee 14-16, 53115 Bonn, FRG}
\date{\today}
\maketitle

\begin{abstract}
  We present a covariant scheme to calculate mesonic transitions in
  the framework of the Salpeter equation for $q\bar{q}$-states. The
  full Bethe Salpeter amplitudes are reconstructed from equal time
  amplitudes which were obtained in a previous paper\cite{Mue} by
  solving the Salpeter equation for a confining plus an instanton
  induced interaction. This method is applied to calculate
  electromagnetic form factors and decay widths of low lying
  pseudoscalar and vector mesons including predictions for CEBAF
  experiments. We also describe the momentum transfer dependence for
  the processes $\pi^0,\eta,\eta'\rightarrow\gamma\gamma^*$.
\end{abstract} \pacs{}

\narrowtext
\section{Introduction} \label{I}
In two previous papers \cite{Mue,RMMP} we presented a quark model for
light mesons based on the Salpeter equation.  We investigated a kernel
that incorporates confinement and a residual instanton induced quark
interaction\cite{Hoo,SVZ,Bla}, which in this framework leads to the
correct masses and flavor mixing of the \(\pi\) and \(\eta\) mesons.
In general we obtained a satisfactory description of the mass spectrum
of the low lying pseudoscalar and vector mesons.  We also calculated
various decay observables such as the weak decay constants, the
\(\gamma\gamma\)-decay width of the pseudoscalars and the leptonic
widths of vector mesons. A comparison with nonrelativistic results
revealed the relevance of the relativistic treatment (including the
correct normalization of the bound states) for the description of
these observables, especially for the two photon width of the pion.

All these transitions involve a non-hadronic final state and therefore
could be calculated in the rest frame of the bound state where also
the amplitudes were determined.  A relativistic quark model however
should also be able to describe reactions with a mesonic final state.
If we consider e.g. weak decays of heavy to light mesons or
electromagnetic scattering with large momentum transfer, the outgoing
meson will recoil with relativistic velocity. The calculation of such transitions
between mesonic states therefore involves a boost of at least one of
the meson amplitudes. A covariant formulation of the Salpeter equation
\cite{WM} enables to treat this boost correctly and thus to
investigate the region of large momentum transfer which will be
available e.g. in CEBAF experiments in the near future
\cite{CEB1,CEB2,CEB3}.

A second important ingredient of any relativistic quark model is an
adequate treatment of the off-shell properties of the quarks. Especially for
mesonic states with large binding energy the negative energy Dirac-components
become essential.  If one considers form factors at high momentum
transfer, the quarks are highly off shell. The Salpeter model
presented here allows for a consistent inclusion of these effects.

In section \ref{II} of this paper we briefly repeat the covariant
formulation of the Salpeter equation and the calculation of the full
Bethe-Salpeter (BS) amplitude as well as its transformation
properties. In section \ref{III} we sketch the calculation of the
electromagnetic current in the Mandelstam formalism.
Finally we present our results for the electromagnetic form factors for
\mbox{\(0^-\! \rightarrow\! 0^-\)} and \mbox{\(1^-\!  \rightarrow \!
  0^-\)} transitions, the corresponding decay widths and the form
factors for the processes \mbox{\(\pi^0,\eta,\eta' \rightarrow
  \gamma\gamma^*\)}.

\section{The Salpeter equation and the reconstruction of the full
  Bethe Salpeter amplitude} \label{II} 
The Bethe-Salpeter equation for the amplitude
\begin{equation}
  \left[ \chi_P(x) \right]_{\alpha\beta} = \left\langle\, 0 \,
  \left|\, T\,\Psi^1_{\alpha}(\eta_1x)\,\bar{\Psi}^2_{\beta}(-\eta_2x)
  \, \right|\, P \,\right\rangle \label{1} ,\end{equation} in momentum
  space reads \cite{BS,RMMP}:
\begin{eqnarray}
\label{4}
\chi_P(p) &=& S^F_1(p_1)\, \int\! \frac{d^4 p'}{(2\pi)^4}\,
[-i\,K(P,p,p')\,\chi_P(p')]\, S^F_2(-p_2)
\end{eqnarray} 
where \( p_1=\eta_1P+p ,\; p_2=\eta_2P-p\) denote the
momenta of the quark and antiquark respectively, \(P\) is the four
momentum of the bound state and \(S^F\) and K are the Feynman
propagators and the irreducible interaction kernel. Here
\(\eta_1,\,\eta_2\) are two arbitrary real numbers satisfying
\(\eta_1+\eta_2=1\).

The Salpeter equation neglects retardation effects in the interaction
kernel in the rest frame of the bound state. This can be written
covariantly as \(K(P,p,p') = V(p_{\perp},\,p_{\perp}') \label{int}\)
where \mbox{\( p_{\perp} = p-(Pp/P^2)\,P \)} \cite{WM}. Furthermore it
is assumed that the propagators are given by their free form
\(S^F_i(p) = i/(p\!\!  /-m_i+i\epsilon)\) with $m_i$ an effective
constituent quark mass.  Introducing the Salpeter or equal time
amplitude in the rest frame of the bound state
\begin{equation}
  \Phi(\vec{p}\,) := \int\!
  dp^0\,\chi_P(p^0,\vec{p}\,)\,\Big|_{P=(M,\vec{0}\,)}
\end{equation}
one arrives at the well-known Salpeter equation \cite{Sa}, i.e.
\begin{eqnarray}
\Phi(\vec{p}\,) &=& 
\int \!\!\frac{d^3p'}{(2\pi)^3}\,
\frac{\Lambda^-_1(\vec{p}\,)\,\gamma^0\,
[(V(\vec{p},\vec{p}\,')\,\Phi(\vec{p}\,')]
\,\gamma^0\,\Lambda^+_2(-\vec{p}\,)}
{M+\omega_1+\omega_2} 
 \nonumber \\
 &-&
\int \!\!\frac{d^3p'}{(2\pi)^3}\,
\frac{\Lambda^+_1(\vec{p}\,)\,\gamma^0\,
[(V(\vec{p},\vec{p}\,')\,\Phi(\vec{p}\,')]
\,\gamma^0\,\Lambda^-_2(-\vec{p}\,)}
{M-\omega_1-\omega_2}
 \label{9}
\end{eqnarray} with the projectors \(\Lambda^{\pm}_i = (\omega_i \pm
H_i)/(2\omega_i)\), the Dirac Hamiltonian
\(H_i(\vec{p}\,)=\gamma^0(\vec{\gamma}\vec{p}+m_i)\) and
$\omega_i=(m^2_i+\vec{p\,}^2)^{1/2}$ as well as
$V(\vec{p},\vec{p}\,'):=V(p_{\perp},\,p_{\perp}')\Big|_{P=(M,\vec{0}\,)}$.

The amplitudes $\Phi$ have been calculated by solving the Salpeter
equation for an interaction kernel including a confining plus a
residual instanton induced interaction. The parameters have been fixed
in order to reproduce the masses of the pseudoscalar and vector
mesons, the weak decay constant of the pion and the leptonic width of
the $\rho$-meson. The results have been presented in ref.\cite{Mue},
we use model V1 therein for the following calculations. The
confinement interaction has been described by a timelike vector spin
structure
\begin{equation}
  \left[V_C^V(\vec{p},\vec{p}\,')\,\Phi(\vec{p}\,')\right] = -{\cal
    V}_C((\vec{p}-\vec{p}\,')^2)\;
  \gamma^0\,\Phi(\vec{p}\,')\,\gamma^0
\label{conf}
\end{equation} 
as a scalar confinement leads to an
RPA-instability of the Salpeter equation\cite{Arch,Piek}.  The scalar
function \({\cal V}_C\) in coordinate space is given by a linearly
rising potential \({\cal V}_C(r) = a_c+b_c r\) in analogy to
nonrelativistic quark models, see e.g.\cite{GI,Bla}.

In order to reproduce the spectrum of the pseudoscalar mesons, we used
an additional instanton induced interaction given by 't Hooft
\cite{Hoo,SVZ,Bla,Mue}.  It acts only on pseudoscalar and scalar
mesons and has the form
\begin{eqnarray}
  [V_T(\vec{p},\vec{p}\,')\,\Phi(\vec{p}\,')] = 4\,\,G \left[
  1\,\mbox{tr}\,\left(\Phi(\vec{p}\,')\right) +
  \gamma^5\,\mbox{tr}\,\left(\Phi(\vec{p}\,')\,\gamma^5\right)\,\right]
\label{thokern} \end{eqnarray}
where G is a flavor dependent coupling constant. Here summation over
flavor and a regularizing Gaussian function have been suppressed (see
\cite{Mue} for more details).

The calculation of transition matrix elements between bound states
involves the knowledge of the full BS amplitude $\chi_P(p)$ which has
to be reconstructed from the equal time amplitude $\Phi(\vec{p}\,)$.
From the Bethe-Salpeter equation itself one finds that the
amputated BS amplitude or vertex function 
\(
 \Gamma_P(p)  := 
[S^F_1(p_1)]^{-1} \,\chi_P(p)\;[S^F_2(-p_2)]^{-1}
\)
may be computed in the rest frame from the equal time amplitude as
\begin{eqnarray} 
  \Gamma_P(p)\left|_{_{P=(M,\vec{0}\,)}}\right.  \; =\;
\Gamma(\vec{p}\,) \; =\;
  -i\! \int\!\! \frac{d^3p'}{(2\pi)^4}
  \left[ V(\vec{p},\vec{p}\,')\Phi(\vec{p}\,')\right]
\label{vert}
\end{eqnarray} 
From the transformation law for the Dirac field
operators \(U_{\Lambda}\Psi(x)U_{\Lambda}^+ =
S_{\Lambda}^{-1}\Psi(\Lambda x)\) and the corresponding properties for
the bound state \(\left|\, P \,\right\rangle\) with mass \(M\) one
derives for the transformation property of the BS amplitudes under a
Lorentz transformation \(\Lambda\):
\begin{equation}
    \chi_{\Lambda P}^{J\, M_J}(p) = \sum_{M'_J} \,\;        
        S_{\Lambda}^{}\,\;\chi_P^{J\, M'_J}(\Lambda^{-1}p)\,\;
        S_{\Lambda}^{-1} 
        \,\;\;  {D_{M_J M'_J}^{J\; *}} \left(u(\Lambda,P)\right),
\label{trans}
\end{equation} 
where \(u(\Lambda,P) :=\Lambda_{\Lambda
  P}^{-1}\;\Lambda \;\Lambda_{P}^{}\) is the corresponding Wigner
rotation and we defined the boost \(\Lambda_P\) by
\(P=\Lambda_P(M,\vec{0})\).

For a pure boost \(\Lambda_P\) we can thus calculate the full BS
amplitude in any reference frame as
\begin{equation}
  \chi_P(p) = \;
  S_{\Lambda_P}^{}\;\;\chi_{(M,\vec{0})}(\Lambda_P^{-1}p)\;\;
  S_{\Lambda_P}^{-1}.
\label{boo}
\end{equation}

Because of the covariant ansatz of the interaction kernel this
kinematical boost gives the solution of the equation for any momentum
\(\vec{P}\) of the bound state.

\section{Transition Amplitudes in the Salpeter Formalism}
 \label{III}
\begin{figure}[h]
  \centering
  \leavevmode
  \epsfxsize=0.65\textwidth
  \epsffile{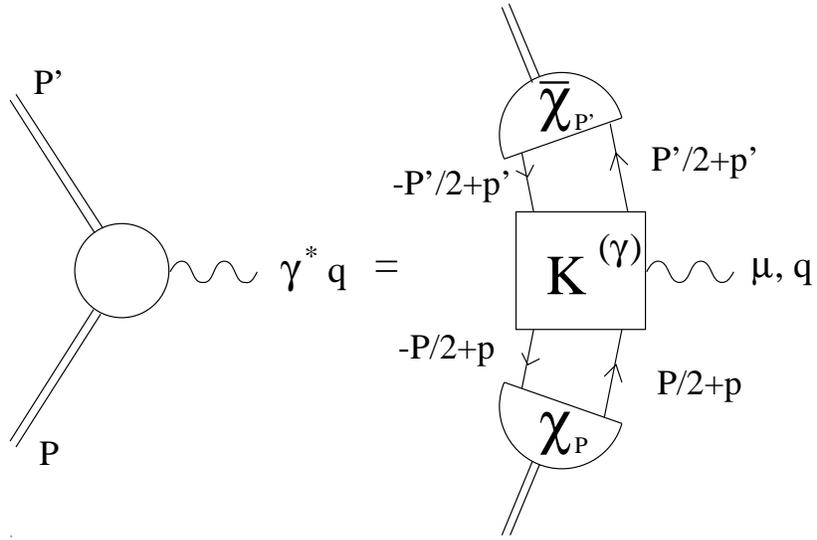}
  \vspace{1cm}
  \caption{The electromagnetic current in the Mandelstam formalism
    calculated from the BS amplitudes \protect{\(\chi_P,
    \bar{\chi}_{P'}\)} and 
    the kernel \protect{\(K^{(\gamma)}\)} and the definition of 
    the relevant momenta.} 
  \label{Kgam2}
\end{figure}

The general prescription for the calculation of any current matrix
element between bound states has been given by Mandelstam \cite{Ma}, see
e.g. \cite{Lu} for a textbook treatment. Consider for example the
electromagnetic current operator: it may be calculated from the BS
amplitudes and a kernel \(K^{(\gamma)}\) as shown in Fig.\ref{Kgam2}.
\(K^{(\gamma)}\) denotes a kernel irreducible with respect to the
incoming and outgoing quark antiquark pair, i.e. it includes all
diagrams that may not be divided by just cutting the quark and the
antiquark line.

\begin{figure}[h]
  \centering
  \leavevmode
  \epsfxsize=0.75\textwidth
  \epsffile{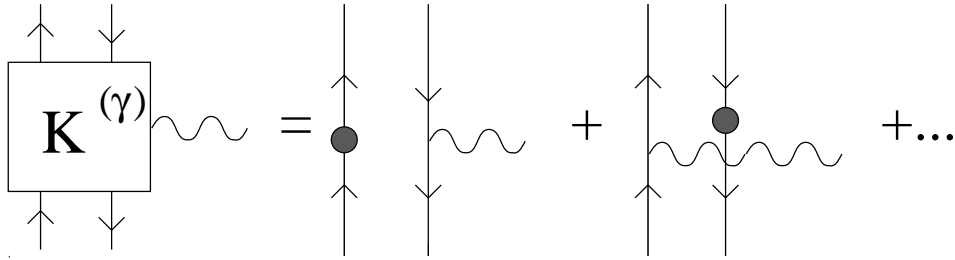}
  \vspace{1cm}
  \caption{Perturbative expansion of the kernel
    \protect{\(K^{(\gamma)}\)} in lowest order of the strong
    interaction; the full circle denotes an inverse quark
    propagator.}
  \label{Kgampt}
\end{figure}

In lowest order the kernel shown in Fig.\ref{Kgampt} reads explicitly
\begin{eqnarray}
  \lefteqn{K^{(\gamma)}_{\mu}(P,q,p,p') =} \\ & & -
  e_1\gamma_{\mu}^{(1)}\,{S^F_2}^{-1}(-P/2+p)\,\delta(p'-p+q/2)\; -
  e_2\gamma_{\mu}^{(2)}\,{S^F_1}^{-1}( P/2+p)\,\delta(p'-p-q/2)
  \nonumber
\end{eqnarray} where \(p\) and \(p'\) denote the relative momenta of
the incoming and outgoing \(q\bar{q}\) pair, $e_1$ and $-e_2$ are the
charge of the quark and antiquark, \(q=P-P'\) is the momentum transfer
of the photon and we use without loss of generality
\(\eta_1=\eta_2=1/2\), as the result is independent of this choice.

The Dirac coupling to pointlike particles is consistent with
the use of free quark propagators. As one of the tasks of
this work is to investigate whether the various properties of the low
lying mesons may be described in terms of constituent
quarks, we neglect their internal structure in the present treatment.
The Ward identity for the free two particle propagator is thus
trivially fulfilled.

For equal mesons in the initial and final state and zero photon
momentum this is consistent with the general normalization condition
for bound states as given by Cutkosky \cite{Cu}, which we already used
in \cite{Mue,RMMP}. In this way the form factor is properly normalized.

For the electromagnetic current coupling e.g. to the first quark we
have explicitly:
\begin{eqnarray}
  \lefteqn{\left\langle\,P'\,\left|\,j_{\mu}^{(1)}(0)\,
  \right|\,P\,\right\rangle =} \\ &= & -
e_1\;\int\!\!\frac{d^4p}{(2\pi)^4}\;\; tr \; \left\{
\bar{\Gamma}_{P'}(p-q/2)\;{S^F_1}(P/2+p-q)\;
\gamma_{\mu}\;{S^F_1}(P/2+p)\;\Gamma_{P}(p)\;{S^F_2}(-P/2+p)\right\}
\nonumber
\end{eqnarray} 

With the transformation properties given in eq.(\ref{trans}) one can
show that the current matrix elements {\em transform covariantly} --
the evaluation thus may be performed in any reference frame.  The
actual calculation of the current is done in the rest frame of the
incoming meson. According to eq.(\ref{vert}) the vertex function of
the initial meson then does depend only on the spacelike three
momentum \(\vec{p}\) and not on \(p^0\).

For \(Q^2=0\) and elastic transitions we have \(P'=P=(M,\vec{0})\), so
that also the outgoing vertex function \(\bar{\Gamma}=
-\gamma_0\Gamma^+\gamma_0\)\cite{RMMP} does depend only on
\(\vec{p}\). The only \(p^0\) dependence is thus contained in the one
particle propagators so that the integral may be calculated
analytically by contour integration according to the Feynman
prescription. The remaining integration in \( |\vec{p}|\) is done
numerically.

For space-like momentum transfer \(Q^2>0\) or nonequal mesons the
outgoing vertex function has to be boosted according to eq.(\ref{boo}):
\begin{equation}
  \Gamma_{P'}(p-q/2)\; = \; S_{\Lambda_{P'}}^{}\;\;
  \Gamma_{(M',\vec{0})}(\vec{p}_{out})\;\; S_{\Lambda_{P'}}^{-1}.
\label{boo2}
\end{equation} with \(p_{out}= \Lambda_{P'}^{-1}(p-q/2)\). This
vertex function thus also depends on the zero component
\(p^0\) of the relative momentum of the incoming \(q\bar{q}\) pair,
although it has no singularities on the real axis. This means that we
cannot close the contour in the \(p^0\)-plane, as we don't know the
analytic structure of \(\Gamma\). 
We therefore perform the $p^0$ integration by principal value
technique. The real part of the form factor is given by sum of the residues
of the six poles of the one particle propagators via 
\begin{equation}
  i\,\int_a^b \!\!dp^0\, \frac{f(p^0)}{p^0-p^0_k \pm i\epsilon} = 
  \pm \, \pi\,
  f(p^0_k) \;+\;i\,\int_a^b \!\!\!\!\!\!\!\!
  P\, dp^0 \,\frac{f(p^0)}{p^0-p^0_k} 
\end{equation} 
for an isolated pole and
\begin{equation}
  i\,\int_a^b \!\!dp^0\, \frac{f(p^0)}{(p^0-p^0_k \pm i\epsilon)^2} 
  = \pm \, \pi\,
  f'(p^0_k) \;+\;i\,\int_a^b \!\!\!\!\!\!\!\!
  P\, dp^0 \,\frac{f(p^0)-f(p^0_k)}{(p^0-p^0_k)^2} 
\end{equation} 
for a double pole and $a\,<\,p^0_k\,<b$.  Here $f(p^0)$
is a real function and the phase $i$ comes from the product
of the one particle propagators. This summation over all the
one-particle poles means that we take into account the positive {\em
  and} negative energy component of the amplitude, which is important
for reactions involving relativistically bound states or large
momentum transfer.  The imaginary part vanishes due to the hermiticity
of the current and time reversal invariance.

As our model includes confinement, we also have mesons with mass \(M\)
larger than the sum \(m_1 + m_2\) of the constituents. The pinching
singularities, that appear in general for such states for values of
the relative momentum $\vec{p}$ where both particle and antiparticle
are on mass shell, are canceled by the zeros of the corresponding
trace of the spin part, so that the integral remains well defined.
This is due to the fact that the projection of the vertex function on
positive energies $\Gamma_{pos} (\vec{p}\,) :=\Lambda^+_1(\vec{p}\,)
\Gamma(\vec{p}\,) \Lambda^-_2(-\vec{p}\,)$ vanishes if both the quark
and the antiquark are on shell. This means that the decay amplitude of
the meson bound states into a free quark and antiquark vanishes so
that confinement in this channel is guaranteed.  The remaining
integration in \(|\vec{p}|\) and \(\cos \Theta_p\) is done
numerically.

The spin part of the current is evaluated by a standard trace
technique appropriate for the particle antiparticle formalism.

The radial part of the vertex functions as well as the Salpeter
amplitudes have been expanded in a basis of eleven Laguerre functions.
The results are found to be stable within a large range of the
scale parameter of the basis.

\section{Results and Discussion}
As already mentioned in sec.\ref{II}, the parameters of the model were
adjusted in a previous work \cite{Mue} (model V1) to reproduce the
mass spectrum and the decay observables with non-hadronic final states
of the pseudoscalar and vector mesons (we refrained from readjusting
them in order not to loose predictive power). The electromagnetic
transitions calculated below thus have {\em no free parameters}, and
give a further test of the Salpeter model for mesons as well as 
predictions for future experiments.

\subsection{The Form Factors \protect{\(M\rightarrow M'\; \gamma^*\)}}
\label{A}
\begin{figure}[h]
  \centering
  \leavevmode
  \epsfxsize=1.0\textwidth
  \epsffile{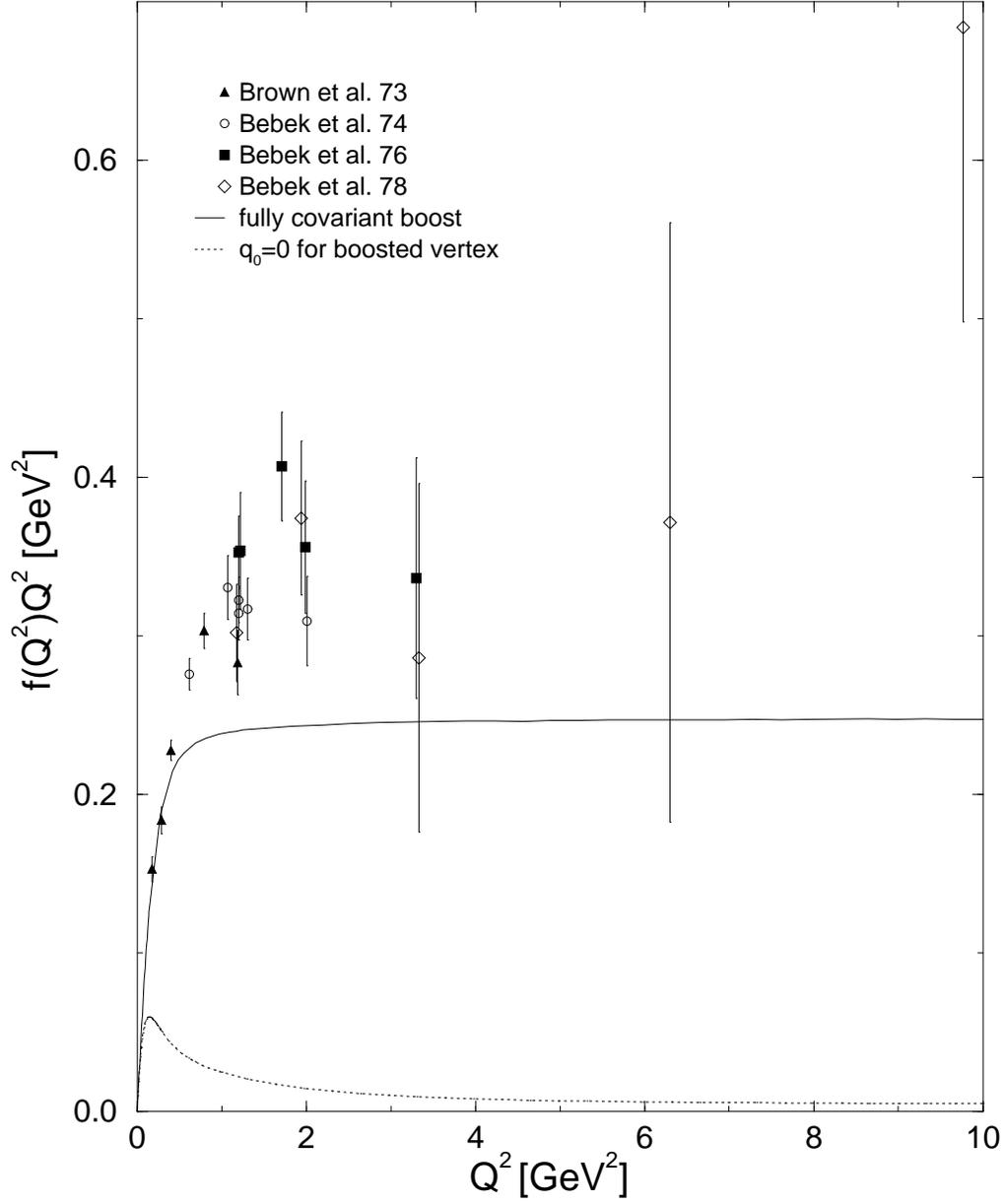}
  \caption{The pion form factor for large momentum transfer compared
    to results from pion photoproduction, see \protect{\cite{Beb}} and
    the references given therein. The solid curve represents the
    calculation with the correct boost of the vertex function, the
    dashed line is obtained, if the zero component of its relative
    momentum is neglected}
  \label{ffpi}
\end{figure}

The electromagnetic form factor \(f(Q^2)\) of pseudoscalar mesons is
defined by
\begin{equation}
\left\langle\,P'\,\left|\,j_{\mu}(0)\,
            \right|\,P\,\right\rangle  \; =\; e
            \;f(Q^2)\, (P'+P)_{\mu}  
\end{equation} 
with $e$ the total electric charge of the meson.  Consider first the
pion which in a constituent quark model is the most deeply bound
state.  In Fig.\ref{ffpi} we compare our results (full line) for
\(Q^2\cdot f(Q^2)\) with experimental data up to \(10 \,
\mbox{GeV}^2\) \cite{Beb}.  The agreement is rather good even for high
momentum transfer and shows that \(f(Q^2)\) behaves as \(1/Q^2\) for
large \(Q^2\) (the error bars for some of the data points are still
very large, so that new CEBAF experiments \cite{CEB1} for this process
are interesting). It supports the hypothesis of Isgur and Llewellyn
Smith\cite{isgur} stating that the form factor in this region should
be explained by nonperturbative effects.  We also would like to
mention similar calculations in the quasi-potential formalism by
Tiemeijer and Tjon \cite{tjon} and in a separable ansatz including
chiral symmetry breaking by Ito, Buck and Gross \cite{IBG1}.  Their
calculations show a stronger fall off for higher momentum transfer.
To analyze this effect we performed a calculation where we neglected
the $p^0$ dependence of the outgoing vertex function (dashed line).
Obviously the inclusion of this fourth component is essential for the
correct treatment at high momentum transfer. A covariant treatment of
the {\em four} relative coordinates of the BS amplitude thus is
mandatory for such processes.

However for small momentum transfer of the order of the quark mass we
find that the form factor of the pion is not strictly monotonic and
depends sensitively on the form of the interaction kernel, so that the
determination of the charge radius becomes ambiguous. This strong
dependence is a result of the large binding energy of the pion and
shows the limits of the Salpeter formalism for such states, at least
for models that include a confinement interaction.

\begin{figure}[h]
  \centering
  \leavevmode
  \epsfxsize=1.0\textwidth
  \epsffile{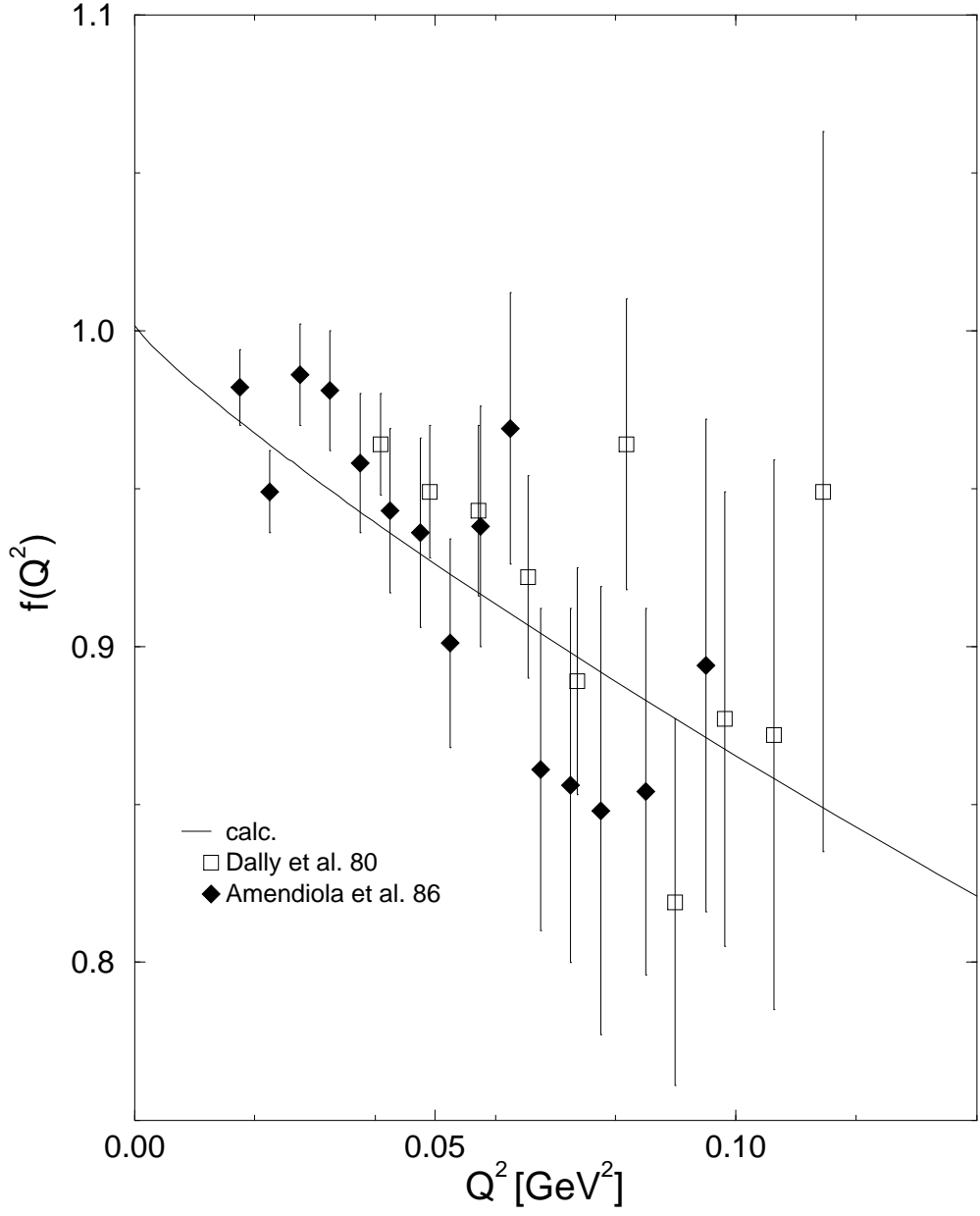}
  \caption{The charged kaon form factor for small momentum transfer
    data from electron scattering\protect{\cite{Ame,Dal}}}
  \label{ffk}
\end{figure}

\begin{figure}[h]
  \centering
  \leavevmode
  \epsfxsize=1.0\textwidth
  \epsffile{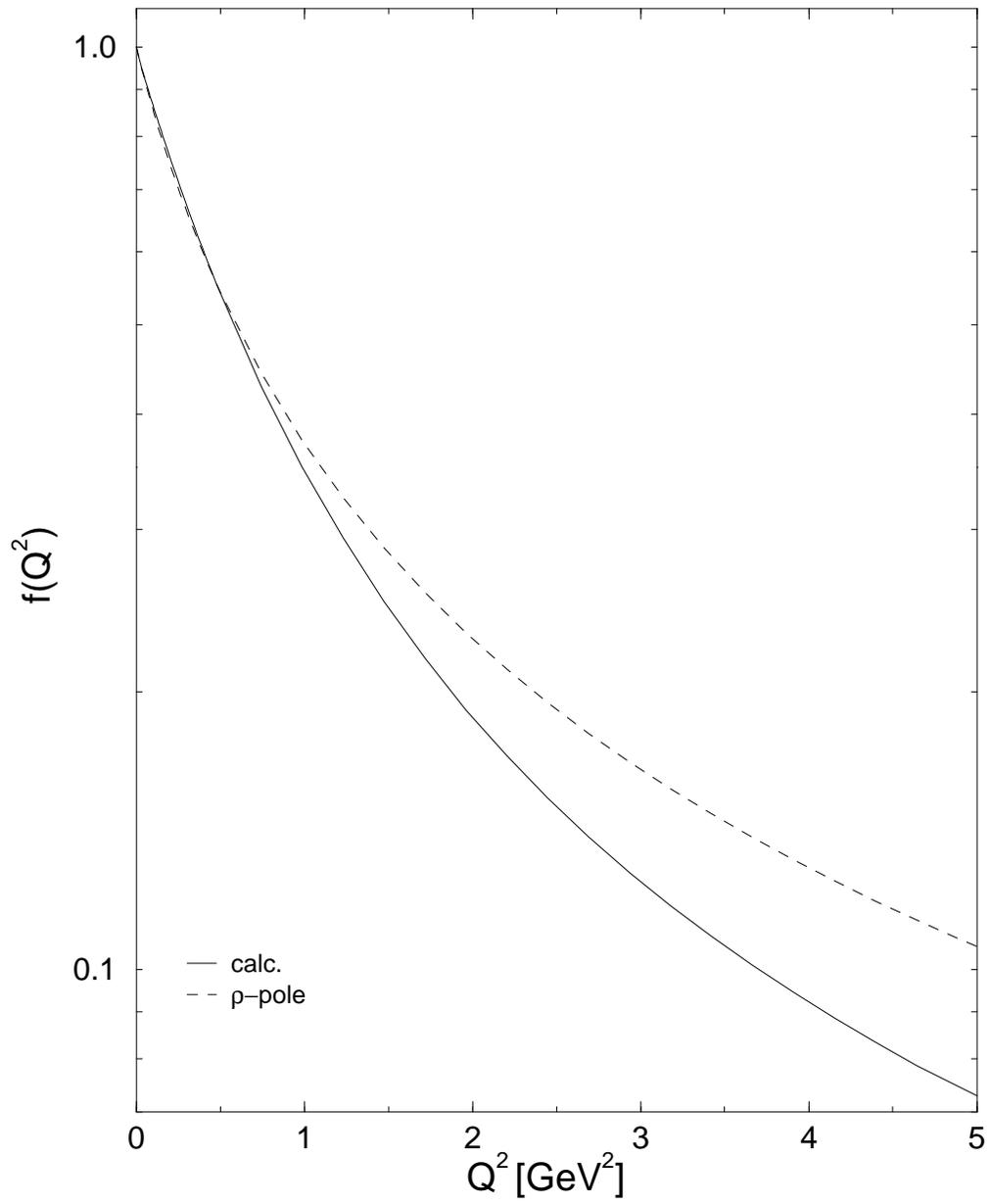}
  \caption{The charged kaon form factor for large momentum transfer
    (solid line) and comparison to a $\rho$-pole motivated by VDM
    (dashed line)}
  \label{ffkhigh}
\end{figure}

In the case of the kaon we obtain a very good description of the form
factor at small momentum transfer, see Fig.\ref{ffk}. We find an
electromagnetic charge radius for the charged kaon of
\mbox{\(<r^2_{K^{\pm}}> ^{1/2}_{calc}= 0.60\,\mbox{fm}\)} as compared
to \mbox{\(<r^2_{K^{\pm}}> ^{1/2}_{exp}= 0.58\pm0.04\,
  \mbox{fm}\)}\cite{Ame} or \mbox{\(<r^2_{K^{\pm}}> ^{1/2}_{exp}=
  0.53\pm0.05\, \mbox{fm}\)}\cite{Dal} from electron scattering data.
For the neutral kaon we obtain \mbox{\(<r^2_{K^0}>_{calc} =-0.070\,
  \mbox{fm}^2\)} as compared to \mbox{\(<r^2_{K^0}>_{exp}= -(0.054\pm
  0.026) \, \mbox{fm} ^2\)}\cite{Mol}.  In view of these agreements it
is interesting to test the charged form factor in the region of high
momentum transfer, which will be accessible in a CEBAF
experiment\cite{CEB2}.  Our prediction is plotted in Fig.\ref{ffkhigh}
and compared to a vector dominance model (VDM). A deviation from a
simple $\rho$-monopole ansatz $f(Q^2)=1/(1+Q^2/M_{\rho}^2)$ is
predicted to appear for \(Q^2>1\,\mbox{GeV}^2\).

We studied the effects of the relativistic treatment by calculating
only those contributions, where the quarks have positive energy. Apart
from the noncovariance of the calculation this would correspond to the
use of a reduced Salpeter equation. We find that the contribution of
the negative energy states to the charged kaon form factor (and
therefore to the normalization) is $25\%$ for zero momentum transfer.
It gives a radius of \mbox{\(<r^2 _{K^{\pm}}>^{1/2}_{pos.\,
    energy}=0.67\,\mbox{fm}\)},
which is off the experimental result.  Relativistic effects thus
play an important role for light mesons even at small $Q^2$.

The form factor $f_{\rho\pi}(Q^2)$ for the process \(\rho\rightarrow
\pi\gamma^*\) (or \(\omega\rightarrow \pi\gamma^*\) ) may be defined
according to
\begin{equation}
\left\langle\,\pi(P')\,\left|\,j_{\mu}(0)\,
            \right|\,\rho(P,\,\lambda)\right\rangle  \; =\;e\;
            \;\frac{f_{\rho\pi}(Q^2)} {M_{\rho}}
            \varepsilon_{\mu\nu\sigma\tau}
            \;\epsilon^{\nu}(P,\lambda)\;P'^{\sigma}\;
            P^{\tau}
\end{equation} 
where $\epsilon(P,\lambda)$ denotes the polarization vector of the
$\rho$(or $\omega$)-meson with spin projection $\lambda$.

\begin{figure}[h]
  \centering
  \leavevmode
  \epsfxsize=1.0\textwidth
  \epsffile{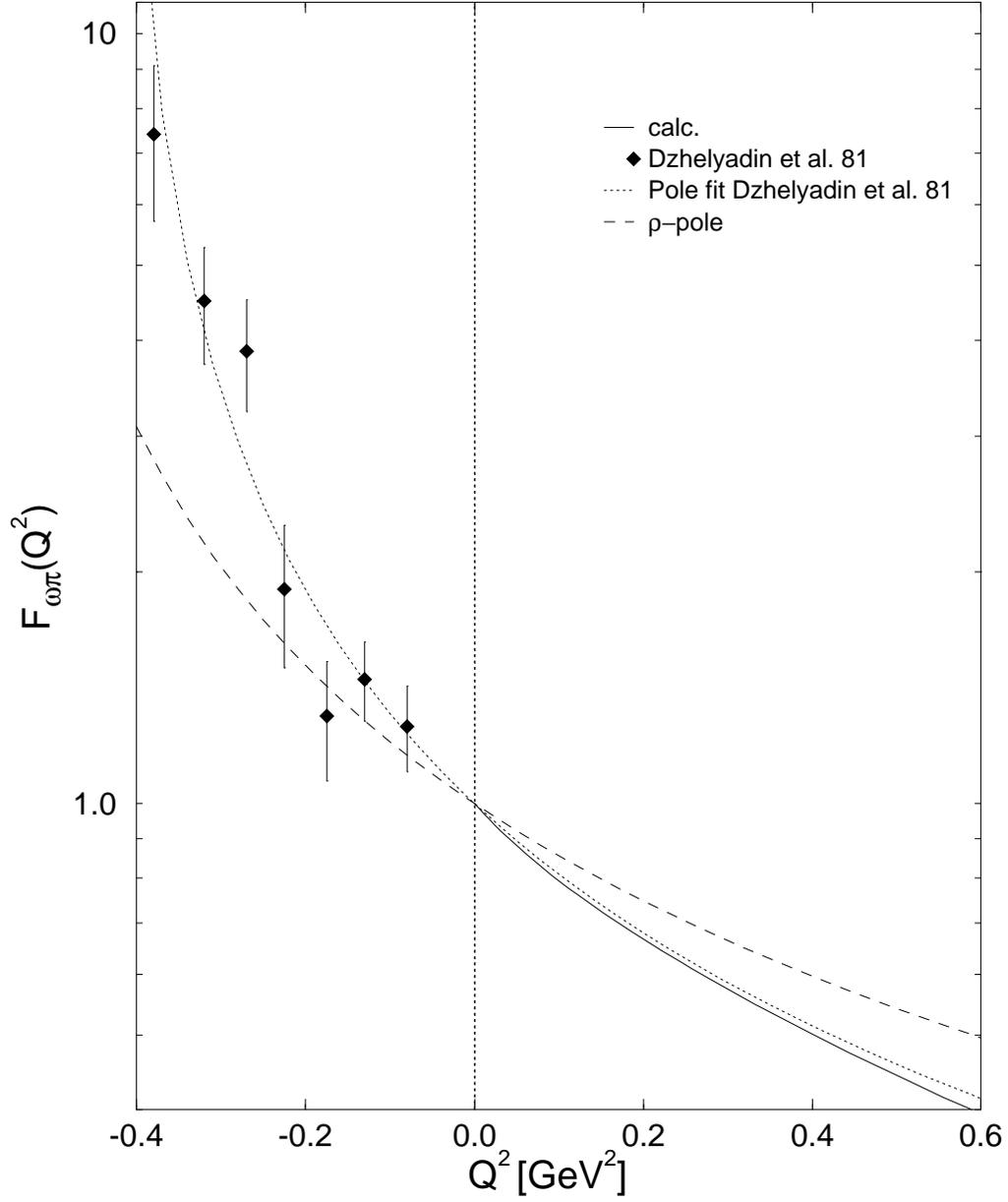}
  \caption{Comparison of the normalized \(\omega\pi\gamma^*\) form
    factor (solid line) in the space-like region with an extrapolation
    of experimental data in the time-like region
    \protect{\cite{Dzhel}} (dotted line) and with a $\rho$-pole ansatz
    motivated by vector dominance (dashed line)}
  \label{ffomegapi}
\end{figure}

The transition \mbox{$\omega\rightarrow\pi\gamma^*$} has been measured
in the time like region via the decay \mbox{$\omega \rightarrow
  \pi^0\mu^+\mu^-$}\cite{Dzhel}, where the normalized quantity
\(F_{\omega\pi}(Q^2) :=f_{\omega\pi}(Q^2) /f_{\omega\pi}(0)\) is
fitted by a simple pole ansatz \(F_{\omega\pi}(Q^2)
=1/(1+Q^2/\Lambda_{\omega\pi}^2)\) with $\Lambda _{\omega\pi} ^{exp} =
(0.65\pm0.03) \,\mbox{GeV}$. We compare the experimental results and
fit to our calculation in the space like region\footnote{a calculation
  of the quantity $F_{\omega\pi}(Q^2)$ becomes meaningless in the
  timelike region, as our model does not guarantee confinement in the
  $\rho\rightarrow\pi\, q\bar{q}$ channel, so that the graph diverges
  for \mbox{$Q^2 \rightarrow -(m_q+m_{\bar{q}})^2$}} in
Fig.\ref{ffomegapi}.  Our curve would correspond on this scale to
$\Lambda_{\omega\pi} ^{calc} = 0.63\,\mbox{GeV}$.  The extrapolation
of the data to the space like region and our prediction thus agree
excellently, especially if we compare to a simple $\rho$-pole
motivated by vector dominance, i.e.  $\Lambda_{\omega\pi} ^{VDM} =
0.77\,\mbox{GeV}$, which is far off the experimental data. Thus we
have found a process, where a relativistic quark model is superior to
the phenomenological vector dominance model even at small momentum
transfer.

\begin{figure}[h]
  \centering
  \leavevmode
  \epsfxsize=1.0\textwidth
  \epsffile{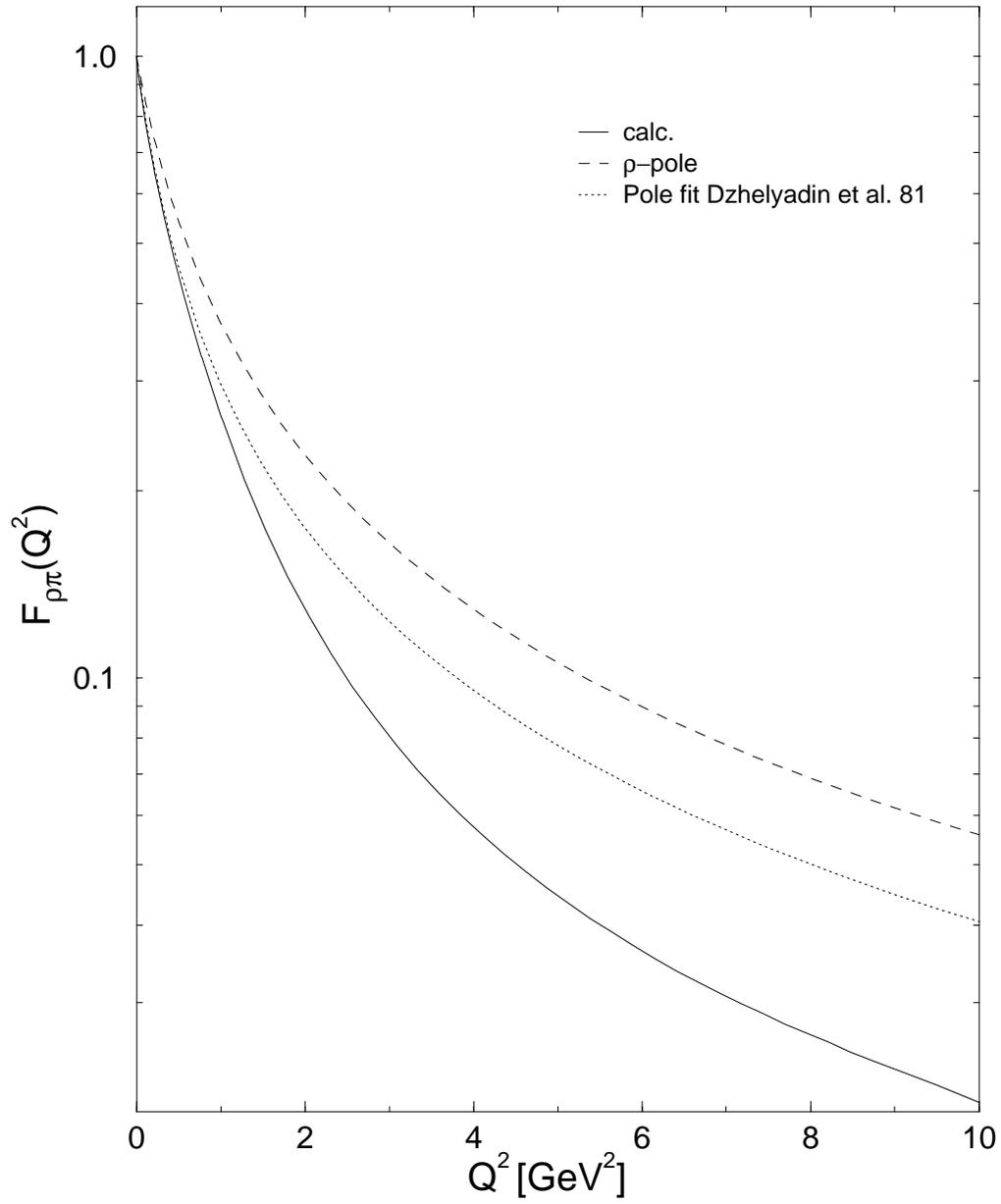}
  \caption{The normalized \(\rho\pi\gamma^*\; (\omega\pi\gamma^*)\) 
    form factor for large momentum transfer}
  \label{ffrpg}
\end{figure}

The \(\rho\rightarrow \pi\gamma^*\) form factor, which in our model is
degenerate with \(\omega\rightarrow \pi\gamma^*\), is particularly
interesting for the calculation of meson exchange currents of the
deuteron form factor. As there is also experimental interest in this
quantity\cite{CEB3,CEB4}, we plot our prediction for large momentum
transfer in Fig.\ref{ffrpg} and compare it to a simple $\rho$-pole and
to the pole fit of \cite{Dzhel} for $\omega\rightarrow\pi^0\gamma^*\)
discussed above\footnote{assuming SU(2) flavor symmetry for
  \(F_{\omega\pi}\) and \(F_{\rho\pi}\) which of course must not be
  true experimentally}. At momentum transfer larger than
\(1\,\mbox{GeV}^2\), which is particularly important for relativistic
calculations of the deuteron, we find a deviation even from the latter
one. The calculation by Ito et al.\cite{IBG1} obtained similar results
in this context for \(\rho\rightarrow \pi\gamma^*\). These authors
also discussed contributions beyond the impulse approximation
including an interaction current\cite{riska}, however neglecting the
confinement problem.  Our absolute value of $f_{\rho\pi}(0)$ is less
accurate and will be discussed in the context of the electromagnetic
decay widths.

\begin{figure}[h]
  \centering
  \leavevmode
  \epsfxsize=1.0\textwidth
  \epsffile{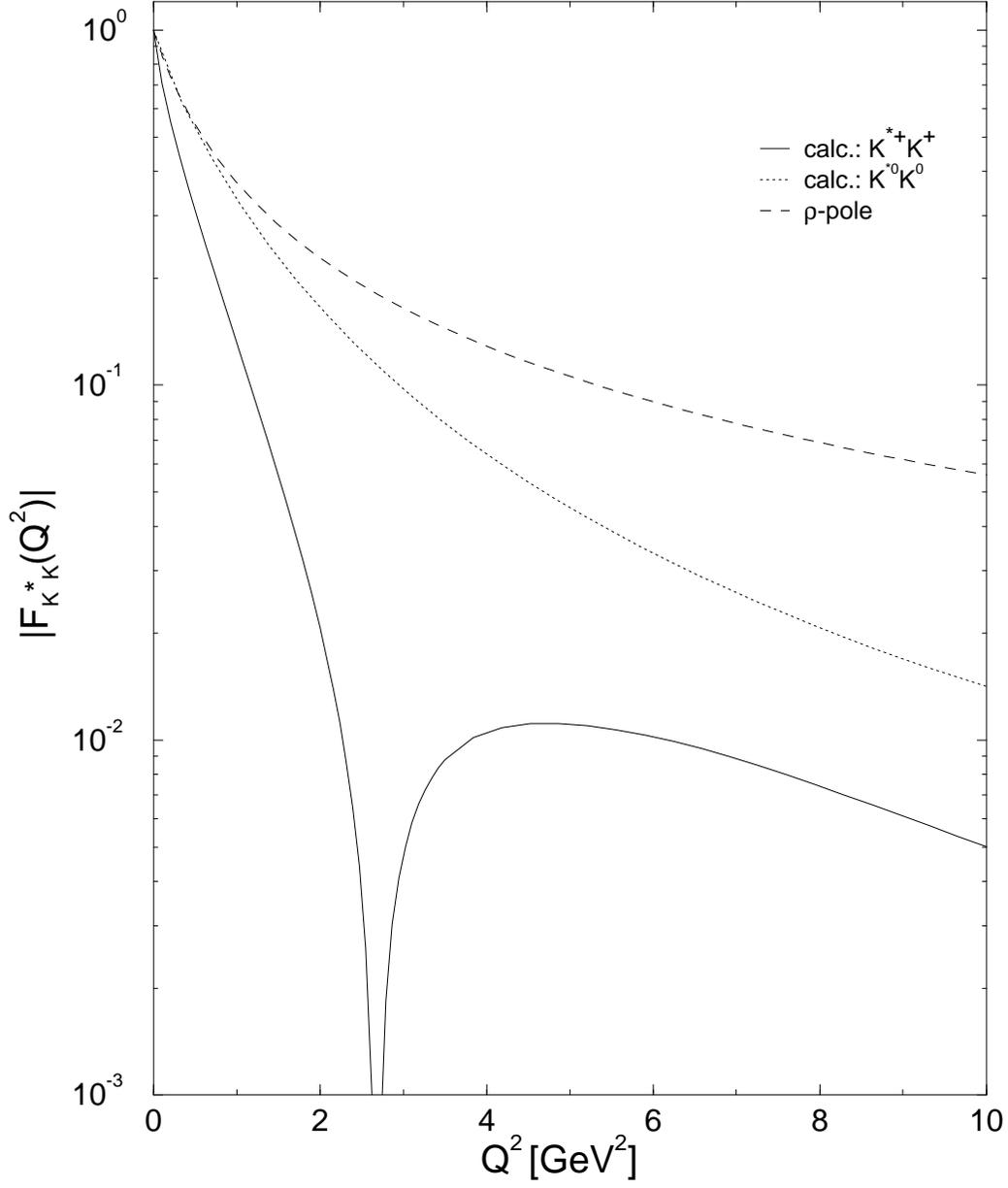}
  \caption{The normalized neutral (dotted line) and charged (solid line) 
    \(K^*K\gamma^*\) form factors compared to VDM (dashed line)}
  \label{ffkstk}
\end{figure}

The corresponding form factors in the strange sector
$K^{*\pm}\rightarrow K^{\pm}\gamma^*$ and $K^{*0}\rightarrow
K^{0}\gamma^*$ are extremely interesting quantities, as the currents
coupling to the quark and the antiquark differ because of their
unequal masses.  Our results in Fig.\ref{ffkstk} show for the neutral
process at least at small momentum transfer a nearly VDM-like
behavior. However in the charged case the picture is totally
different: the negative interference between the two currents leads to
a zero in the form factor at $Q^2_0 = 2.7\,\mbox{GeV}^2$, a region
which is already highly relativistic. The effect may be understood
qualitatively in a VDM type model, where the coupling to the quark and
antiquark is assumed to be proportional to their respective magnetic
moments and to a propagator of the corresponding vector meson $\rho$
or $\Phi$. The result however depends sensitively on the ratio of the
mass of the strange ($m_s$) and nonstrange ($m_n$) constituent quark.
We varied these masses keeping the sum of them as well as the other
parameters fixed and obtained a dependence on the ratio $m_s/m_n$ that
is listed in Tab.\ref{paramn}. Our original fit in \cite{Mue} used
$m_n=170\,\mbox{MeV}$ and $m_s=390\,\mbox{MeV}$, i.e. corresponds to a
ratio $m_s/m_n=2.3$.

Because of the accurate results for the corresponding decay widths
(see next section) we consider these calculations even more reliable
than in the $\rho\pi\gamma$ case.  In view of the sensitivity of the
zero in the form factor we therefore would encourage an experimental
investigation of this interesting phenomenon e.g. at the CEBAF
facility, thus providing empirical information on the ratio of the
constituent quark masses.

\subsection{The decay width \protect{\(M\rightarrow M'\; \gamma\)} for
  the ground state mesons} 
There exist several measurements of decay processes of an excited
meson to a state with lower mass by emission of a single real
photon\cite{PDG}. They provide a suitable test of the BS amplitudes
especially for resonances where no detailed study of form factors is
available. The results for the transitions between vector and
pseudoscalar mesons are summarized in Tab.\ref{emdecays}. If the mass
difference is large, the final meson is emitted with relativistic
velocity, so that a covariant framework is essential.

In the semirelativistic ansatz of Godfrey and Isgur \cite{GI} the
nonrelativistic decay formulae have been modified by terms $(m/E)^n$
with m the quark mass and E its energy, which however forces to use
new ad hoc parameters n. The relativistic framework
presented here includes these effects automatically.

The widths for decays with a pion in the final state are generally
underestimated by a factor of two\footnote{from SU(2)-isospin symmetry
  the decay width $\rho^0\rightarrow\pi^0\gamma$ should be the same as
  for $\rho^{\pm}\rightarrow\pi^{\pm}\gamma$ and a factor nine smaller
  than the width for $\omega\rightarrow\pi^0\gamma$ and therefore this
  experimental value has to be considered with care}. Our results are
consistent with a calculation of Tiemeijer\cite{tiem} in a similar
equal time formalism.  This again indicates that the Salpeter
formalism is not fully satisfactory in the case of such deeply bound
states as the pion.

The transitions of strange mesons $K^*\rightarrow K\gamma$ are in
excellent agreement with measurements for both the neutral and the
charged channel, consistent with the good results of the kaon form
factor.  The kaon therefore seems to be well understood in the
Salpeter model. Again the width for the charged decay does sensitively
depend on the ratio of the constituent quark masses, see
Tab.\ref{paramn}, giving a ratio of $m_s/m_n = 1.5-1.9$, whereas the
neutral decay is almost stable and does not restrict this quantity.

The electromagnetic decays involving a $\eta$ ($\eta'$) meson in the
final (initial) state are a sensitive test of the $n\bar{n}$- and
$s\bar{s}$-component of their BS amplitude and therefore of the
interaction that induces the flavor mixing.  The decays
$\rho,\omega\rightarrow\eta\gamma$ involve the
$n\bar{n}$-component, $\Phi\rightarrow\eta\gamma$ the
$s\bar{s}$-component.  The agreement is excellent for all the three
values reconfirming the good description of the mixing coefficients
for the $\eta$ given in \cite{Mue}.  For the $\eta'$ decays into
$\rho$ and $\omega$ we find an overestimation of a factor of around
two consistent with the fact that the $n\bar{n}$-component of this
meson is too large in our model compared to the semi-empirical
value extracted from the $J/\Psi$-decay\cite{Mue}.

Our prediction for the decay width $\Phi\rightarrow\eta'\gamma$
includes the estimated error from the inaccuracy in the calculated
meson masses which enters the transition matrix elements.  As we
underestimate the $s\bar{s}$-component of the $\eta'$ only by a small
amount\cite{Mue}, we expect our result for this experimental value to be 
quite accurate.

\subsection{The Form Factors \protect{\(M\rightarrow \gamma \gamma^*\)}}
\begin{figure}[h]
  \centering
  \leavevmode
  \epsfxsize=0.65\textwidth
  \epsffile{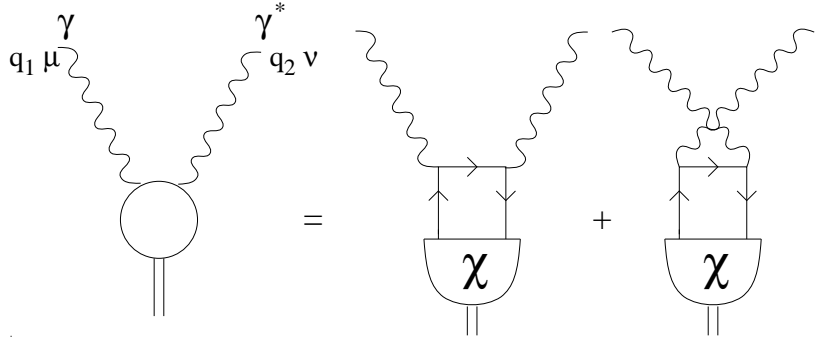}
  \caption{Feynman graph for the decay into two photons}
  \label{gamgam}
\end{figure}

The structure of the BS amplitude for neutral pseudoscalar mesons may
be tested by the production via a virtual and an (almost) real photon
as done at 
\(\gamma\gamma\) facilities of \(e^+e^-\) colliders \cite{TPC,Cello}.
In lowest order the process is given by the graphs in
Fig.\ref{gamgam}. If one of the photons is on shell, i.e.  $q_1^2=0$,
the amplitude may be parameterized as
\begin{equation}
  T_{\mu\nu}(q_1,q_2) =
  \epsilon_{\mu\nu\alpha\beta}\,q_1^{\alpha}\,q_2^{\beta}\,f_{M\gamma}
  (Q_2^2)          
\end{equation}
Referring to the parameterization of the experimental data given in 
\cite{TPC,Cello} we define a width 
\begin{equation}
    \Gamma(Q^2) = \frac{M^3}{64\pi}f^2_{M\gamma}(Q^2)
\end{equation}
which for \(Q^2=0\) gives the decay width for a pseudoscalar into two
real photons.

\begin{figure}[h]
  \centering
  \leavevmode
  \epsfxsize=0.75\textwidth
  \epsffile{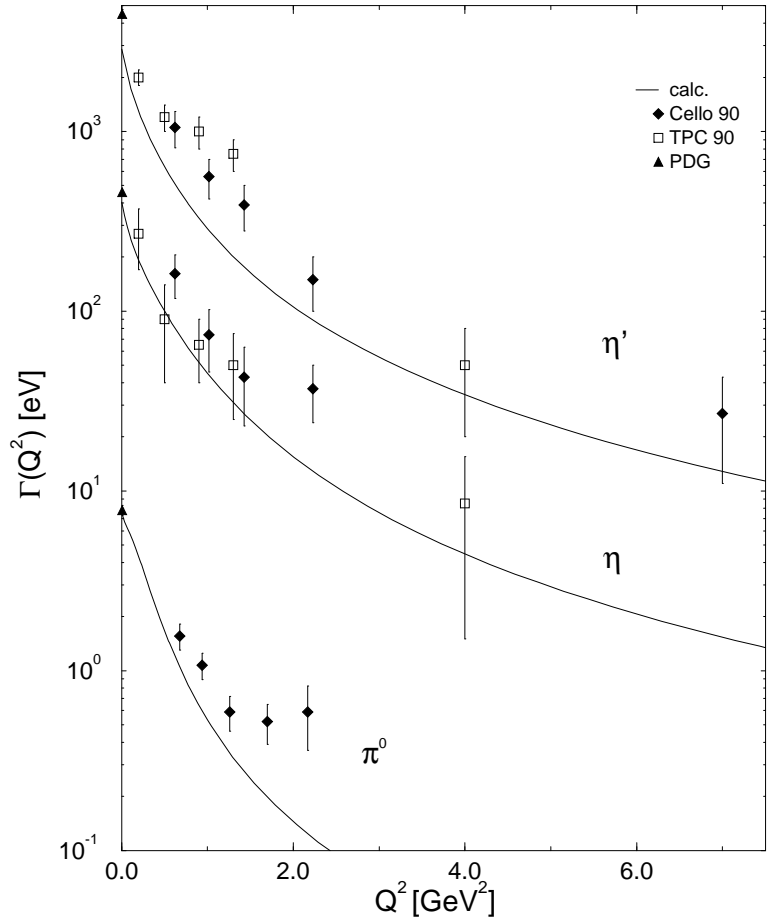}
  \caption{The \protect{\(\pi^0,\eta\) and \(\eta' \rightarrow
      \gamma\gamma^*\)} form factors compared to experimental
    data\protect{\cite{TPC,Cello}}}  
  \label{peegg}
\end{figure}

In Fig.\ref{peegg} we have plotted the results for \(\Gamma(Q^2)\) for
the \(\pi^0,\eta\) and \(\eta'\) including the experimental
results\cite{TPC,Cello}.  The decay widths have already been published
and are in good agreement with experimental data\cite{Mue}.

The width for the process \(\pi^0\rightarrow \gamma\gamma^*\) depends
sensitively on the quark mass, a result that has also been found by
Ito et al.\cite{IBG1}. Their optimal value of the nonstrange quark
mass \(m_n=250\, \mbox{MeV}\) is in rough agreement with our result of
\(m_n=170\, \mbox{MeV}\), which had been adjusted in order to obtain the
correct pion decay constant\footnote{a larger quark mass would give a
  smaller width at zero momentum transfer, but also a smaller slope}.
However as in the case of the charged pion form factor we find an
unnatural structure for $\Gamma_{\pi^0\rightarrow \gamma\gamma}(Q^2)$
for low momentum transfer due to the strong binding of the pion. Also
in this process the width for larger momentum transfer is somewhat
underestimated.

The structure of the \(\eta\) form factor can be described almost
quantitatively up to \(4\,\mbox{GeV}^2\), which underlines the good
description of this meson for the observables discussed above.
Although the $\eta'$ width is too small by about 30\%, the dependence
on \(Q^2\) is well reproduced up to \(8\,\mbox{GeV}^2\).

\section{Summary and Conclusion}
Starting from a relativistic quark model based on the Salpeter
equation that includes confinement and an instanton induced flavor
mixing interaction, we investigated the electromagnetic properties of
the light pseudoscalar and vector mesons including the isoscalar
states. In order not to loose predictive power we used the BS
amplitudes of a former work\cite{Mue}, all the present results thus
were calculated with no additional free parameter.

Especially for the $\eta$ and K meson, but also for the lowest vector
mesons we find an excellent description of all available observables.
The flavor mixing for the $\eta$, which can be measured in the decays
$\rho,\omega,\Phi\rightarrow\eta\gamma$, is correctly reproduced. The
kaon form factor for small momentum transfer as well as the $K^*
\rightarrow K\gamma$-widths are in almost quantitative agreement.  The
phenomenological extrapolation of the $\omega \rightarrow \pi\gamma^*$
form factor from the data in the time like region agrees with our
quark model prediction in the space like region, but differs
significantly from the standard vector dominance model. We present
predictions for the processes $\rho\rightarrow\pi\gamma^*$ and the
kaon form factor in the large $Q^2$ regime which will be measured at
CEBAF in near future. We found that the transition form factor
$K^{*\pm} \rightarrow K^{\pm}\gamma^*$ represents an interesting
observable, as its form depends strongly on the ratio of strange and
nonstrange quark mass.  Because of the negative interference of the
current coupling to quark and antiquark one obtains a zero of the
amplitude, which we predict at $Q^2_0\approx2.7 \mbox{GeV}^2$. From
the decay width into a real photon we find $m_s/m_n=1.9-1.5$.

The description of the $1/Q^2$-behaviour of the charged pion form
factor in the region of high momentum transfer is possible only if the
BS amplitude is boosted correctly, i.e. if the full dependence of the
relative four momentum is taken into account.  However we reach the
limits of the Salpeter ansatz in the case of the pion due to its
strong binding. We find that the charged and neutral form factor on
the scale of the quark mass become extremely sensitive to the
interaction -- in our model there are structures not apparent in
experimental data. We could not find a kernel that includes
confinement and is able to describe the form factors in this region.

The results show that a relativistic treatment of constituent quarks
in the framework of the Salpeter model for mesons including a
relativistic normalization and covariant boosting of the amplitudes is
able to describe almost quantitatively the various properties of the
ground state pseudoscalar and vector mesons. In view of this success
we will apply the formalism to a detailed study of the complete meson
spectrum including one-gluon-exchange for heavy systems, which will be
done in a future work.

{\bf Acknowledgments:} We appreciate the help of S.\ Hainzl, G.\ 
Z\"oller and M.\ Fuchs in numerical calculations as well as
encouraging remarks by K.\ Althoff. One of the authors (C.R.M) would
like to thank the CEBAF theory group for hospitality and interesting
and fruitful discussions. This work was partially supported by the
Bundesministerium f\"ur Forschung und Technologie.

\begin{table}
  \caption{Dependence of the $K$- and $K^*$-mass, weak decay
    constant $f_K$, charged kaon radius, decay widths and the zero of
    the form factor for $K^{*\pm}\rightarrow K^\pm \gamma$ on the
    ratio of the strange and nonstrange constituent quark mass
    $m_s/m_n$ (masses and decay constant given in MeV)}
  \label{paramn}
  \centering
   \begin{tabular}{cccccccc}
       $m_s/m_n$  & $M_K$  & $M_K^*$ 
        & $f_K$ & $<r^2_{K^{\pm}}>$[fm] & $\Gamma_{K^{*\pm}\rightarrow
         K^{\pm}\gamma}\,[\mbox{keV}]$  &  $\Gamma_{K^{*0}\rightarrow
         K^{0}\gamma}\,[\mbox{keV}]$  & $Q^2_0\,[\mbox{GeV}^2]$     \\
     \hline
         3.0 & 535 & 895 & 179 & 0.61 & 78  & 114 & 2.1  \\
         2.3 & 510 & 880 & 183 & 0.60 & 64  & 112 & 2.7   \\
         1.8 & 485 & 865 & 185 & 0.59 & 52  & 111 & 4.8   \\
         1.5 & 475 & 860 & 187 & 0.59 & 45  & 110 & $>$10   \\
         1.0 & 465 & 855 & 189 & 0.57 & 27  & 109 & --   \\
         Experiment & 495 & 892 & 164 & 0.58$\pm$0.04& 50$\pm$5 & 
                                               117$\pm$10 &  \\       
\end{tabular}
\end{table}

\begin{table}
  \caption{Comparison of experimental and calculated electromagnetic 
    meson decay widths}
  \label{emdecays}
  \centering
   \begin{tabular}{cccccc}
       Mesonic decay width [\mbox{keV}]  &  experimental \cite{PDG} 
                                                      & calculated\\
     \hline
          \(\Gamma(\rho^{\pm} \rightarrow \pi^{\pm}\gamma)\)
                  & 68 \(\pm\) 7     & 38     \\
          \(\Gamma(\rho^{0} \rightarrow \pi^{0}\gamma)\)
                  & 121 \(\pm\) 31     & 38     \\
          \(\Gamma(\omega \rightarrow \pi\gamma)\)
                  & 717 \(\pm\) 43     & 335     \\
          \(\Gamma({K^*}^{\pm} \rightarrow K^{\pm}\gamma)\)
                  & 50 \(\pm\) 5     & 64     \\
          \(\Gamma({K^*}^{0} \rightarrow K^{0}\gamma)\)
                  & 117 \(\pm\) 10     & 112     \\
          \(\Gamma(\rho \rightarrow \eta\gamma)\)
                  & 58 \(\pm\) 10     & 50     \\
          \(\Gamma(\omega \rightarrow \eta\gamma)\)
                  & 4.0 \(\pm\) 1.7     & 5.6     \\
          \(\Gamma(\phi \rightarrow \eta\gamma)\)
                  & 56.9 \(\pm\) 2.9     & 60     \\
          \(\Gamma(\eta' \rightarrow \omega\gamma)\)
                  & 5.9 \(\pm\) 0.9     & 12.7     \\
          \(\Gamma(\eta' \rightarrow \rho\gamma)\)
                  & 59 \(\pm\) 6     & 122     \\
          \(\Gamma(\phi \rightarrow \eta'\gamma)\)
                  & \(<\) 1.8     & 0.18\(\pm\)0.02     \\
\end{tabular}
\end{table}

\newpage

\end{document}